# Methods for Accelerating Conway's Doomsday Algorithm (part 1)

Chamberlain Fong [spectralfft@yahoo.com]
06/19/2010

**Abstract**

We propose a modification of a key component in the Doomsday Algorithm for calculating the day of the week of any given date. In particular, we propose to replace the calculation of the required expression:

$$\left\lfloor \frac{x}{12} \right\rfloor + x \bmod 12 + \left\lfloor \frac{x \bmod 12}{4} \right\rfloor$$

with

$$2y + 10\,(y \bmod 2) + z + \left\lfloor \frac{2\,(y \bmod 2) + z}{4} \right\rfloor$$

where x is an input 2-digit year;
y is the tens digit of x;
z is the ones digit of x;

We argue the fact that our modification operates on individual base-10 digits makes the algorithm easier to calculate mentally.

**Keywords:** Doomsday Algorithm, Doomsday Rule, Calendar Algorithm, John Horton Conway, Mental Calculation

**Introduction**

The Doomsday algorithm (or Doomsday rule) is an algorithm for calculating the day of the week for any given calendar date. It was invented by the eminent mathematician, John Horton Conway. It was designed to be simple enough that, with practice, one can do the calculation in his head without paper or pencil and come up with an answer in just a few seconds. Conway himself was able to do the algorithm calculations in his head in about 2 seconds[1].

For brevity, we won't discuss the full algorithm here. Instead we refer the readers to excellent write-ups[2][3] that describe the algorithm in full detail. We will, however, summarize key components of the algorithm to point out some important definitions and nomenclature to be used in this paper.

The Doomsday algorithm's input is a calendar date of the form MM/DD/YYYY where MM is the month, DD is the day, and YYYY is the year. YYYY can further be broken down to its constituent century CC and year within the century YY. The output of the algorithm is a number between 0 to 6 that corresponds to each of the 7 days fo the week.

The key equation of the Doomsday algorithm can be described a sum (modulo 7) of three terms:

day_of_the_week = ( doomscentury + doomsyear + doomsmonth ) mod 7

where:

>doomscentury(CC) is a function of the input date's century
>doomsyear(YY) is a function of the input date's 2-digit year within a century
>doomsmonth(MM,DD) is a function of the input date's calendar month and day.

In this paper, we only really care about the doomsyear term. We will only gloss over the doomsmonth and doomscentury terms. The doomsmonth term is what takes advantage of the fact that April 4, June 6, August 8, October 10, and December 12 all fall on the same day of the week for any calendar year. The doomscentury term accounts for the quirk in the Gregorian calendar system where years that are divisible by 100 are not leap years except when they are divisible by 400. We will not attempt to simplify or discuss further the doomscentury and doomsmonth terms for the rest of this paper.

The doomsyear formula provided by Conway[4][5] is:

$$\left\lfloor \frac{x}{12} \right\rfloor + x \bmod 12 + \left\lfloor \frac{x \bmod 12}{4} \right\rfloor$$

where x is the input date's 2-digit year within a century. The addition is always modulo 7, so the resulting sum is a number between 0 and 6, inclusive.

Conway actually got this expression from Lewis Carroll who did early work on perpetual calendar algorithms[6][7]. In this paper, we shall refer to use of this expression as the *Carrollian method* for calculating doomsyear.

**Related Work**

In practice, the Carrollian method for doomsyear calculation is a bottleneck of the Doomsday algorithm. Aside from having to do divisions by 12 and 4, one has to keep in memory 3 numbers to add. The Carrollian method is, actually, already an acceleration method[2] for the true doomsyear value that needs to be computed:

$$x + \left\lfloor \frac{x}{4} \right\rfloor$$

which, in practice, is even more difficult to calculate mentally for two-digit input numbers; especially when followed by a required modulo 7 operator. Several people have suggested other acceleration methods for calculating doomsyear. Conway himself offered a lookup table acceleration for it[3][8]. His acceleration method is arguably the fastest but it requires the memorization of 18 numbers and extra rules for handling leap years. In 2008, Mike Walters[9] came up with an elegant method for calculating doomsyear that involves addition of multiples of 11, a division by 2, and a negation modulo 7 operator. Bob Goddard[10] made refinements to this method in 2009. Our proposed acceleration method is just as simple as Mike Walters' and Bob Goddard's method. In addition, our proposed acceleration method is amenable to a table-lookup memorization similar to Conway's lookup table acceleration.

In part 2 of this paper, we will present a significant enhancement to Mike Walters' 2008 acceleration method. This new method does not require any divisibility-by-4 tests to calculate doomsyear. Ultimately, we present the reader with two different methods for simplified calculation of the doomsyear term in parts 1 and 2 of this paper.

**Simplified Formula**

Let us break down the two-digit year x into its constituent digits y and z, where y is the tens digit, and z is ones digit. That is,

$$y = \left\lfloor \frac{x}{10} \right\rfloor$$

$$z = x \bmod 10$$

For example, if x = 74, then y= 7 and z = 4. If x = 88, then y= 8 and z= 8.

Having defined y and z in terms of x, we propose the replacement doomsyear function as

```
doomsyear(y,z) =  2 y + 10(y mod 2) + z + leaps
```

where (*y mod 2)* is really just a decision function to tell whether y is odd or even.

(y mod 2) = 1 if y is odd
(y mod 2) = 0 if y is even

We define an extra variable called *leaps* as the number of leap years between the start of the y decade and the z year. If the start of a decade is a leap year, we don't count it. But if the year z is a leap year, we do include it. For example, if x = 88, the decade starts at 80 and we have *leaps* = 2 because 84 and 88 are leap years. If x = 74, the decade starts at 70 and we have *leaps* = 1 because 72 is a leap year. In general, the variable *leaps* can only have three values: 0, or 1, or 2. There can't be more than 2 leap years after the start of a decade. Remember, we never include the start of the decade in our leap count.

The explicit formula for *leaps* is

$$leaps = \left\lfloor \frac{2\,(y \bmod 2) + z}{4} \right\rfloor$$

But, in practice, it is easier to count the leap years of a decade than to evaluate this expression mentally.

In summary, our proposed method for calculating doomsyear breaks the 2-digit input year into its constituent digits; and does simple arithmetic on these digits. Afterwards, there is the addition of a leap year correction term. Our method essentially requires only 3 additions and a multiplication by 2. Moreover, since the input is broken down to single digit numbers, the calculations done only involve small numbers and are easy to do mentally. In fact, our doomsyear arithmetic only involves positive integers in between 0 to 39. This range of numbers happens to be close to the "calendar range" of numbers which cannot be avoided in the calculations of any calendar algorithm.

There are no explicit divisions or subtractions in our doomsyear formula. However, there are implicit divisions when one is figuring out the leap year correction term and when one decides whether the decade is odd or even. There are also implicit subtractions when one is doing a modulo 7 operation afterwards. Nevertheless, all doomsyear calculation methods require these implicit divisions and subtractions!

**Examples**

Let's calculate the doomsyear term for these years:

1) 1974:  y = 7, z = 4

   doomsyear = 2*7 + 10*1 + 4 + leaps  =  14 + 10 +  4 + 1 = 29 = **1**   (mod 7)

   leaps = 1 because 1972 is a leap year

2) 2040:  y = 4, z= 0

   doomsyear = 2*4 + 10*0 + 0 + leaps = 8 + 0 + 0 + 0  =  **1**    (mod 7)

   leaps = 0

3) 2010: y = 1, z = 0

   doomsyear = 2*1 + 10*1 + 0 + leaps = 2 + 10 + 0 + 0 =12 =  **5**   (mod 7)

   leaps = 0

4) 1988: y = 8, z= 8

   doomsyear = 2*8 + 10*0 + 8 + leaps  =  16 + 0 + 8 + 2  = 26 =  **5**    (mod 7)

   leaps = 2  because 1984 and 1988 are leap years

5) 2007:  y = 0, z= 7

   doomsyear = 2*0 + 10*0 + 7 + leaps  =  0 + 0 + 7 + 1  =  8  =  **1**   (mod 7)

   leaps = 1 because 2004 is a leap year

6) 1998:  y = 9, z=8

   doomsyear = 2*9 + 10*1 + 8 + leaps = 18 + 10 + 8 + 2 =  38 = **3**   (mod 7)

   leaps = 2 because 1992 and 1996 are leap years

**Tips and Tricks**

<u>Memorization and Lookup Tables</u>
Let's define the *decade anchor* to be the *2y + 10(y mod 2)* subexpression of our doomsyear term. This subexpression only depends on the y decade of the input year.

$$\text{decade\_anchor}(y) = (\, 2y + 10\,(y \bmod 2)\, ) \bmod 7$$

By virtue of memorizing the decade anchor for each decade, one can significantly speed-up the calculation of the doomsyear term.  This memorization is, of course, optional but recommended for speed.  Here is the table to memorize:

| y | decade | 2y + 10 (y mod 2) | decade anchor |
|---|--------|-------------------|---------------|
| 0 | 00's | 0 | **0** |
| 1 | 10's | 12 | **5** |
| 2 | 20's | 4 | **4** |
| 3 | 30's | 16 | **2** |
| 4 | 40's | 8 | **1** |
| 5 | 50's | 20 | **6** |
| 6 | 60's | 12 | **5** |
| 7 | 70's | 24 | **3** |
| 8 | 80's | 16 | **2** |
| 9 | 90's | 28 | **0** |

Table 1: Decade anchor lookup

If one memorizes this simple table of 10 numbers, one can avoid calculating the decade anchor *2y + 10 (y mod 2)* component of the doomsyear formula. The doomsyear formula is thus:

$$doomsyear(y,z) = decade\_anchor(y) + z + leaps$$

An important mnemonic that can aid in memorizing this table involves the observation of its *clamping effect*. That is, for the starting and ending decades of the 00's and the 90's, the decade anchor is 0. Thus, one only really needs to memorize 8 numbers when using this decade anchor lookup table acceleration trick.

Another mnemonic of note is that the 50's decade has anchor 6; and the 60's decade has anchor 5. In effect, 5 maps to 6 ; and 6 maps 5. So there is a crisscross swap near the middle of the table. We call this observation as the *midriff* mnemonic

Look Before You Leap

There are several tricks to make the calculation of the *leaps* term easier. Let's look at a table of possible values for the leap term depending on the digit year z:

| digit year z | leaps | leaps if y is even | leaps if y is odd |
|---|---|---|---|
| 0 | 0 | 0 | 0 |
| 1 | 0 | 0 | 0 |
| 2 | 0 or 1 | 0 | 1 |
| 3 | 0 or 1 | 0 | 1 |
| 4 | 1 | 1 | 1 |
| 5 | 1 | 1 | 1 |
| 6 | 1 or 2 | 1 | 2 |
| 7 | 1 or 2 | 1 | 2 |
| 8 | 2 | 2 | 2 |
| 9 | 2 | 2 | 2 |

Table 2: Possible Values for leap

One can easily see that for digit years 0 and 1, the leaps value is always 0. Likewise, for the digit years 8 and 9, the leaps value is always 2; and for the digit years 4 and 5, the leaps value is always 1. So, we only really need to do leap year calculations when the digit year is 2, 3, 6, or 7. Furthermore, if one checks whether the decade is odd or even, leap year determination can be totally avoided at the expense of more memorization. For a more thorough coverage of divisibility by 4 tricks and leap years, see YingKing Yu's article[11].

Lucky Number 7

Another tip for the calculation of the doomsyear term involves the observation that 7 is equal to 0 in modulo 7 arithmetic. So the z term in the doomsyear formula can be dropped altogether for the years 07, 17, 27, 37, 47 etc.

**Proof?**

We offer no mathematical proof that our doomsyear method is correct. Instead, we intend to show the next best thing. We tabulated all the numbers from 00 to 99 in a spreadsheet and calculated the doomsyear value of each number using the Carrollian method as well as using our new formula. The results are shown in Table 3. They matched exactly on every input number. This is the "brute force" way of proving that the two methods are equivalent. For a rigorous proof of our method, see YingKing Yu's article [11].

**Conclusion**

We presented an acceleration method for calculating the doomsyear term of the Doomsday algorithm. Our method is simple because it essentially only involves additions and a multiplication by 2. Furthermore, we provided further tips and tricks that can give additional speed to our method. In the appendix section of this paper, we show that our acceleration method is similar in form to Conway's lookup table acceleration method, but requires less memorization.

**Acknowledgements**

Thanks to the following people who replied to email inquiries about the Doomsday algorithm: Mike Walters, YingKing Yu, Bob Goddard, Sidney Graham, and Rudy Limeback.

| year | Conway's doomsyear | 2y + 10(y mod 2) + z + leaps | year | Conway's doomsyear | 2y + 10(y mod 2) + z + leaps |
|---|---|---|---|---|---|
| 0 | 0 | 0 | 56 | 0 | 0 |
| 1 | 1 | 1 | 57 | 1 | 1 |
| 2 | 2 | 2 | 58 | 2 | 2 |
| 3 | 3 | 3 | 59 | 3 | 3 |
| 4 | 5 | 5 | 60 | 5 | 5 |
| 5 | 6 | 6 | 61 | 6 | 6 |
| 6 | 0 | 0 | 62 | 0 | 0 |
| 7 | 1 | 1 | 63 | 1 | 1 |
| 8 | 3 | 3 | 64 | 3 | 3 |
| 9 | 4 | 4 | 65 | 4 | 4 |
| 10 | 5 | 5 | 66 | 5 | 5 |
| 11 | 6 | 6 | 67 | 6 | 6 |
| 12 | 1 | 1 | 68 | 1 | 1 |
| 13 | 2 | 2 | 69 | 2 | 2 |
| 14 | 3 | 3 | 70 | 3 | 3 |
| 15 | 4 | 4 | 71 | 4 | 4 |
| 16 | 6 | 6 | 72 | 6 | 6 |
| 17 | 0 | 0 | 73 | 0 | 0 |
| 18 | 1 | 1 | 74 | 1 | 1 |
| 19 | 2 | 2 | 75 | 2 | 2 |
| 20 | 4 | 4 | 76 | 4 | 4 |
| 21 | 5 | 5 | 77 | 5 | 5 |
| 22 | 6 | 6 | 78 | 6 | 6 |
| 23 | 0 | 0 | 79 | 0 | 0 |
| 24 | 2 | 2 | 80 | 2 | 2 |
| 25 | 3 | 3 | 81 | 3 | 3 |
| 26 | 4 | 4 | 82 | 4 | 4 |
| 27 | 5 | 5 | 83 | 5 | 5 |
| 28 | 0 | 0 | 84 | 0 | 0 |
| 29 | 1 | 1 | 85 | 1 | 1 |
| 30 | 2 | 2 | 86 | 2 | 2 |
| 31 | 3 | 3 | 87 | 3 | 3 |
| 32 | 5 | 5 | 88 | 5 | 5 |
| 33 | 6 | 6 | 89 | 6 | 6 |
| 34 | 0 | 0 | 90 | 0 | 0 |
| 35 | 1 | 1 | 91 | 1 | 1 |
| 36 | 3 | 3 | 92 | 3 | 3 |
| 37 | 4 | 4 | 93 | 4 | 4 |
| 38 | 5 | 5 | 94 | 5 | 5 |
| 39 | 6 | 6 | 95 | 6 | 6 |
| 40 | 1 | 1 | 96 | 1 | 1 |
| 41 | 2 | 2 | 97 | 2 | 2 |
| 42 | 3 | 3 | 98 | 3 | 3 |
| 43 | 4 | 4 | 99 | 4 | 4 |
| 44 | 6 | 6 | | | |
| 45 | 0 | 0 | | | |
| 46 | 1 | 1 | | | |
| 47 | 2 | 2 | | | |
| 48 | 4 | 4 | | | |
| 49 | 5 | 5 | | | |
| 50 | 6 | 6 | | | |
| 51 | 0 | 0 | | | |
| 52 | 2 | 2 | | | |
| 53 | 3 | 3 | | | |
| 54 | 4 | 4 | | | |
| 55 | 5 | 5 | | | |

Table 3: Doomsyear values from 00 to 99

**Appendix 1: Some Doomsyear Properties**

The true formula for doomsyear is:

$$doomsyear(x) = x + \left\lfloor \frac{x}{4} \right\rfloor$$

A list of its values from 0 to 99 is shown in table 3. Let us observe some properties of the doomsyear term.

*Property#1 Non-Leap-Year Increment*

We can see from table 3 that the doomsyear increases by 1 for every non-leap year. This follows from the fact that 1 = 365 (mod 7); and the day of the week increases by one after each non-leap year.

*Property#2 Leap-Year Increment*

We can also see from table 3 that the doomsyear increases by 2 for every leap year. This follows from the fact that 2 = 366 (mod 7); and the day of the week increases by two after each leap year.

*Property#3 Periodicity*

The days of the week repeat with a period of 7. The leap years within a century repeat with a period of 4. It is no surprise that doomsyear values repeat every 28 years. 28 is, of course, the least common multiple of 7 and 4.

*Property#4 Quasi-periodicity*

We can observe from table 3 that the doomsyear value almost repeats a count of 0 to 6 every 5 or 6 years except for regular jumps occurring every leap year. We can call this property as quasi-periodicity.

**Appendix 2: Conway's Look-up Table Acceleration Method**

John Horton Conway devised an acceleration method to speed-up the calculation of the doomsyear term. In practice, Conway's method is probably the fastest acceleration method for such, but it involves memorizing 18 numbers and some non-intuitive rules. In this section, we will describe Conway's look-up table method. In the next section, we will compare and contrast our method with Conway's acceleration method.

Conway's lookup table method requires memorizing the years of the century where the doomsyear value is zero. Let us call these numbers as *zero-anchor* years. These are:

| 0 | 6 | 11.5 | 17 | 23 | 28 | 34 | 39.5 | 45 |
|---|---|------|----|----|----|----|------|-----|
| 51 | 56 | 62 | 67.5 | 73 | 79 | 84 | 90 | 95.5 |

There's actually the added complication of the *half-numbers*: 11.5, 39.5, 67.5 and 95.5. These half-numbers mean that the preceding year has doomsyear value 6 and the succeeding year has doomsyear value 1. For example, doomsyear(11) = 6, and doomsyear(12) = 1; doomsyear(67) = 6, and doomsyear(68) = 1. These half-numbers occur because of the increment-by-2 property of doomsyear values during leap years. In effect, a doomsyear of value 0 got skipped in the half-number locations.

The gist of Conway's acceleration method is to find the nearest zero-anchor year to your input year. Afterwards, subtract the zero-anchor year from your input year and add a leap year correction to get the doomsyear value  For this section, we will restrict the zero-anchor to be the less than your input year. It is actually possible to select a zero-anchor greater than your input year, but it requires remembering some additional rules when dealing with leap years and half-numbers.   This is tricky and error-prone, so we will not cover it.

Here are the steps of Conway's acceleration method:
1) Select the nearest zero-anchor year less than your input year.
2) Let $z_0$ be the difference between your input year and the selected zero-anchor. Ignore fractional values of half-numbers in zero-anchor years. That is, treat 11.5, 39.5, 67.5, and 95.5 as 11, 39, 67, and 95 respectively in calculating the difference
3) Count the number of leap years between the zero-anchor and your input year. If the zero-anchor year is a leap year, do not include it. On the other hand, if your input year is a leap year, include it in the leap count. Let us denote this count of leap years as **$leap_0$**
4) Add $z_0$ and **$leap_0$** to get the doomsyear value. If the selected zero-anchor is a half-number, subtract 1 from the sum. We denote this as the *anchor adjustment* term needed for half-number zero-anchors.

To sum it all up, Conway's acceleration method can be described by the equation:

$$doomsyear = anchor\_adjustment + z_0 + leap_0$$

Note that the anchor_adjustment term is almost always zero except for half-number years where it has a value of –1.

Examples:

1) 1974:

   zero-anchor is 1973

   $z_0$ = 1974 − 1973 = 1

   $leap_0$ = 0

   doomsyear = 1 + 0 = **1**

2) 2040:

   zero-anchor is 2039.5

   $z_0$ = 2040 − 2039 = 1

   $leap_0$ =1 because 2040 is a leap year

   doomsyear =  -1 + 1 + 1 = **1**

3) 2010:

   zero-anchor is 2006

   $z_0$ = 2010 − 2006 = 4

   $leap_0$ = 1 because 2008 is a leap year

   doomsyear = 4 + 1 = **5**

4) 1988:

   zero-anchor is 1984

   $z_0$ = 1988 − 1984 = 4

   $leap_0$ = 1  because 1988  is a leap year. Remember, we don't count the zero-anchor

   doomsyear = 4 + 1 = **5**

5) 2007:

   zero-anchor is 2006

   $z_0$ = 2007 − 2006 = 1

   $leap_0$ = 0

   doomsyear = 1 + 0  = **1**

6) 1998:

    zero-anchor is 1995.5

    $z_0$ = 1998 – 1995 = 3

    $leap_0$ = 1 because 1996 is a leap year

    doomsyear = -1 + 3 + 1 = **3**

7) 1914:

    zero-anchor is 1911.5

    $z_0$ = 1914 – 1911 = 3

    $leap_0$ = 1 because 1912 is a leap year

    doomsyear = -1 + 3 + 1 = **3**

8) 1972:

    zero-anchor is 1967.5

    $z_0$ = 1972 – 1967 = 5

    $leap_0$ = 2 because 1968 and 1972 are leap years

    doomsyear = -1 + 5 + 2 = **6**

**Appendix 3: A Comparison of Our Acceleration Method with Conway's**

As we mentioned in the tips and tricks section, our method is amenable to lookup table acceleration. In fact, we claim that after this lookup table acceleration, our method is very similar in form to Conway's acceleration method. Let us compare and contrast the doomsyear equation for our method and Conway's method. These are:

$$doomsyear(y,z) = decade\_anchor(y) + z + leaps$$

and

$$doomsyear(x) = anchor\_adjustment(x) + z_0 + leap_0$$

respectively. We have carefully named different terms of the equations to highlight the similarities between both equations. Let us list down these similarities.

1) Both equations are the sum of 3 terms. Each of these terms corresponds with a counterpart in the other method.
2) Both equations use memorization of an anchor year for the speedup. In our method, the starting year of the decade serves as the anchor. In Conway's method, years with doomsyear value of zero are used as the anchor.
3) Both equations use z as the number of years between the input year and the anchor year. We can consider z as the *offset* from the anchor
4) Both equations contain a leap count correction term that counts the number of leap years between the anchor year and the input year.

We now list down differences between the 2 equations and mention some advantages of our method over Conway's method.

1) In our method, z is not computed. It is part of the input. In Conway's method, $z_0$ has to be calculated by subtracting the zero-anchor year from the input year.
2) In our method, one has to memorize 10 digits for the anchoring. In Conway's method, one has to memorize 18 numbers for the anchoring.
3) In our method, the leap count correction term follows a regular pattern for a given decade and is amenable to another speedup via memorization.
4) In our method, one can always fall back to using the *2y + 10 (y mod 2)* calculation if entries in the lookup table are forgotten.

# Appendix 4: The Carrollian Method Revisited

Recall that the Carrollian method's formula for doomsyear is:

$$\left\lfloor \frac{x}{12} \right\rfloor + x \bmod 12 + \left\lfloor \frac{x \bmod 12}{4} \right\rfloor$$

As we previously mentioned, this expression is a calculation simplification for

$$x + \left\lfloor \frac{x}{4} \right\rfloor$$

Let us rename each of the terms as:

$$anchor_{12} = \left\lfloor \frac{x}{12} \right\rfloor$$

$$z_{12} = x \bmod 12$$

$$leap_{12} = \left\lfloor \frac{x \bmod 12}{4} \right\rfloor$$

With these redefinitions, we can represent the Carrollian method expression as:

$$doomsyear(x) = anchor_{12} + z_{12} + leap_{12}$$

which shares the same form as our proposed doomsyear formula and Conway's lookup-table acceleration formula. All 3 doomsyear formulas have analogously corresponding anchors, z offset values, and leap year corrections.

To summarize: The Carrollian method for doomsyear is anchored every dozen years to make the calculations simpler. Conway's lookup-table acceleration method is anchored at values where doomsyear is zero. Our proposed doomsyear method is anchored to every decade.